\documentclass{article}
    
\usepackage[final,nonatbib]{neurips_2022_ml4ps}

\usepackage[sort&compress,numbers]{natbib}

\usepackage[utf8]{inputenc} 
\usepackage[T1]{fontenc}    
\usepackage{hyperref}       
\usepackage{xcolor}
\hypersetup{
    colorlinks,
    linkcolor={red!50!black},
    citecolor={blue!50!black},
    urlcolor={blue!80!blue}
}
\usepackage{url}            
\usepackage{booktabs}       
\usepackage{amsfonts, amssymb}       
\usepackage{nicefrac}       
\usepackage{bm, bbm}
\DeclareRobustCommand{\bbone}{\text{\usefont{U}{bbold}{m}{n}1}}

\usepackage{amsmath}
\usepackage{cleveref}
\usepackage{glossaries}
\usepackage{graphicx}
\usepackage{physics}
\usepackage{siunitx}
\usepackage{xcolor}
\usepackage{tikz}
\usepackage{lipsum}  
\usepackage{wrapfig}

\newacronym{ddpm}{DDPM}{denoising diffusion probabilistic model}
\newacronym{em}{EM}{Euler-Maruyama}
\newacronym{elt}{ELT}{Extremely Large Telescope}
\newacronym{gan}{GAN}{generative adversarial network}
\newacronym{gp}{GP}{Gaussian process}
\newacronym{hst}{HST}{Hubble Space Telescope}
\newacronym{mc}{MC}{Monte Carlo}
\newacronym{pc}{PC}{predictor-corrector}
\newacronym{rim}{RIM}{recurrent inference machine}
\newacronym{sbm}{SBM}{score-based model}
\newacronym{sde}{SDE}{stochastic differential equation}
\newacronym{sie}{SIE}{singular isothermal ellipsoid}
\newacronym{snr}{SNR}{signal-to-noise}
\newacronym{vae}{VAE}{variational autoencoder}
\newacronym{vesde}{VE SDE}{variance-exploding SDE}

\newcommand{\noise}{\sigma_\mathcal{N}}
\DeclareMathOperator*{\normal}{\mathcal{N}}

\title{Posterior samples of source galaxies in strong gravitational lenses with score-based priors}

\author{%
Alexandre Adam$^{1,2,4}$ \quad Adam Coogan$^{1,2,4}$ \quad Nikolay Malkin$^{1,2}$ \quad Ronan Legin$^{1,2,3,4}$   \\
\textbf{Laurence Perreault-Levasseur}$^{1,2,3,4}$ \textbf{Yashar Hezaveh}$^{1,3,4}$ \quad \textbf{Yoshua Bengio}$^{1,2,5}$ \\
$^1$Université de Montréal \quad $^2$Mila \quad $^3$CCA, Flatiron Institute \quad $^4$Ciela \quad $^5$CIFAR AI Chair\\
\texttt{\{alexandre.adam,adam.coogan,ronan.legin,laurence.perreault.levasseur,}\\\texttt{yashar.hezaveh\}@umontreal.ca}\\
\texttt{\{nikolay.malkin,yoshua.bengio\}@mila.quebec}\\
}

\begin{document}

\maketitle

\begin{abstract}
    Inferring accurate posteriors for high-dimensional representations of the brightness of gravitationally-lensed sources is a major challenge, in part due to the difficulties of accurately quantifying the priors. Here, we report the use of a score-based model to encode the prior for the inference of undistorted images of background galaxies. This model is trained on a set of high-resolution images of undistorted galaxies. By adding the likelihood score to the prior score and using a reverse-time stochastic differential equation solver, we obtain samples from the posterior. Our method produces independent posterior samples and models the data almost down to the noise level. We show how the balance between the likelihood and the prior meet our expectations in an experiment with out-of-distribution data.
\end{abstract}

\section{Introduction}

Strong gravitational lensing -- extreme distortions in the images of distant sources by the gravity of foreground lensing galaxies -- is a powerful tool that can be used to probe the fundamental nature of dark matter, infer the expansion rate of the universe, and study the birth and evolution of nascent galaxies \citep{Treu10}. Inferring the spatial distribution of mass in the foreground lens and the spatial distribution of surface brightness in the background source is an essential component of achieving these scientific goals.

In this work, we ask the question: given a noisy image of a distorted source and the distribution of mass in the lensing galaxy, how can we infer the spatial distribution of surface brightness in the background source? The goal is to sample the posterior $p(\vb{x} \mid \vb{y}, \vb{\kappa})$, where $\vb{x}$ are the variables representing the surface brightness distribution in the background source, $\vb{y}$ is the observed data, and $\vb{\kappa}$ are the variables representing the spatial distribution of mass in the lens.

In noisy and low-resolution data, the source can often be well-described by low-dimensional representations like the Sérsic profile \citep{sersic,Spilker16}. Through their functional forms, these representations implicitly impose a strong prior on the surface brightness.
Higher-quality data, however, reveal rich and complex morphologies, which demand more expressive source models, such as a set of pixels \citep{Warren04,Suyu06}, allowing arbitrarily-complex representations at a finite resolution. These methods require priors that limit $\vb{x}$ to physically-plausible configurations to avoid unphysical source reconstructions caused by overfitting to noise. Such priors have typically taken heuristic and simplistic forms to facilitate calculations, such as gradient or curvature penalties \citep{Suyu06}. Other expressive source models apply similar methods on adaptive grids \citep{VegettiKoopmans09,2015MNRAS.452.2940N}, decompose the source as a linear combination of shapelets \citep{Refregier03,birrer} or wavelets \citep{Galan:2020mnn}, or model the source as an approximate Gaussian process \citep{Karchev:2021fro}. However, these priors are inaccurate since sampling from them does not yield galaxy-like images, which can bias lensing inference.

Recent work has explored the use of machine learning to create better source brightness priors for lensing inference, such as variational autoencoders \citep{Coogan20}, recurrent inference machines \citep{Morningstar:2019szx,Morningstar:2018ase,Adam22RIM}, and continuous neural fields \citep{sid}. These methods respectively have trouble accurately representing the prior over galaxy images, only produce maximum a posteriori parameter estimates, and only implicitly define a prior through the choice of a neural network architecture. Also, denoising diffusion probabilistic models \citep{Sohl-Dickstein2015,ho2020ddpm} have been applied to learn priors over galaxies outside the context of lensing inference \citep{astroddpm}.

In this work we use score-based modeling, formulated in terms of stochastic differential equations (SDEs)\vphantom{\gls*{sde}}, to learn a highly-accurate prior over the source surface brightness for lensing analysis. In combination with a likelihood, this allows us to produce source posterior samples of remarkably high quality using \gls*{sde} solvers. A similar approach was adopted in \citet{Remy2022} to produce samples from the posterior of convergence maps from weak lensing data. These samples enable us to assess the significance of reconstructed source features. Our experiments show how our prior is balanced against the likelihood to enforce that reconstructions to look like training set galaxies in the low-\gls*{snr} regime. This represents a significant step towards accurate inference in high-dimensional spaces.

\section{Inference of underconstrained variables with score-based priors}
The data-generating process of strongly-lensed images of background galaxies can be described by the linear equation
\begin{equation}
    \label{eq:inverse problem}
    \mathbf{y} = A \mathbf{x} + \boldsymbol{\eta} \, ,
\end{equation} 
where $\mathbf{x}\in \mathbb{R}^n$ contains pixel intensities of the undistorted background source image, $A \in \mathbb{R}^{m \times n}$ encodes the lensing distortions (and thus a function of $\kappa$), interpolation over $\vb{x}$ and instrumental effects such as a point spread function, and $\boldsymbol{\eta} \in \mathbb{R}^m$ is additive instrumental noise, here assumed to be distributed as $\mathcal{N}(\vb{0}, \noise^2\bbone)$. 
Since we consider the case where $\kappa$ is known, in our notation we drop the dependence on~$A$ (or $\kappa$) and treat it as a known constant. The likelihood for the data given the source image is therefore $p(\vb{y} \mid \vb{x}) = \mathcal{N}(\vb{y} \mid A \vb{x}, \noise^2\bbone)$.

Our aim is to sample the posterior $p(\vb{x} \mid \vb{y})$, which, by Bayes' theorem, is proportional to the product between the likelihood $p(\vb{y} \mid \vb{x})$ and a prior $p(\vb{x})$. Applying the logarithm thus gives
\begin{equation}
    \log p(\mathbf{x} \mid \mathbf{y}) = \log p(\mathbf{y} \mid \mathbf{x}) + \log p(\mathbf{x}) - \log p(\mathbf{y}) \, .
\end{equation}
The prior effectively gives the probability that any image $\vb{x}$ looks like a galaxy in the absence of a lensed observation $\vb{y}$. Recent advances in generative modeling have shown that the score of the prior, $\grad_{\vb{x}} \log p(\vb{x})$, can be accurately learned from training data and sampled from using score-based modeling \citep{Vincent2011,Sohl-Dickstein2015,ho2020ddpm,song2021sde}.
We now summarize how to train a model $s_\theta(\vb{x})$ to approximate $\grad_{\vb{x}} \log p(\vb{x})$ using score matching and our posterior sampling procedure.

\subsection{Score matching}

Score matching \citep{Hyvarinen2005} is the task of training a model,
$\mathbf{s}_\theta(\mathbf{x}): \mathbb{R}^n \rightarrow \mathbb{R}^n$, to match the score of a probability distribution, $\grad_{\mathbf{x}} \log p(\mathbf{x})$. We use denoising score matching (DSM) \citep{Vincent2011,Alain2014}, which lets us learn an implicit distribution by training a network $\vb{s}_\theta(\vb{x})$ to remove Gaussian noise added to i.i.d.~samples from that distribution. We follow previous works \citep{Song2019ncsn,Lim2020ardae,song2021sde} in averaging the DSM loss over various scales $\sigma(t)$, here indexed by a continuous time variable $t \in [0, 1]$, and conditioning the score model on this time index, $s_\theta(\vb{x}, t) = \boldsymbol{\epsilon}_\theta(\vb{x}, t) / \sigma(t)$, where $\boldsymbol{\epsilon}_\theta$ is the neural network. The loss is obtained by sampling uniformly over $t$, perturbing a training sample by adding noise of the corresponding scale, noted by $\sigma(t)\vb{z}$ with $\vb{z}\sim \mathcal{N}(\vb{0}, \bbone)$, and computing the Fisher divergence between the model and the kernel of the perturbation:
\begin{equation}
    \mathcal{L}_\theta = \mathbb{E}_{t \sim \mathcal{U}(0, 1)}\mathbb{E}_{\mathbf{x} \sim \mathcal{D}} \mathbb{E}_{\vb{z} \sim \normal(\vb{0}, \bbone)} \left[ \left\lVert \boldsymbol{\epsilon}_\theta \left( \mathbf{x} + \sigma(t) \, \vb{z} , t \right) + \vb{z} \right\rVert_2^2 \right] \, .
\end{equation}
This loss is designed to address the manifold hypothesis \citep{Alain2014,Song2019ncsn} and is related to denoising diffusion approaches that rely on a variational formulation \citep{bengio2014deep,Sohl-Dickstein2015,goyal2017variational,Song2019ncsn,ho2020ddpm,graikos2022diffusion} to generate data with a fixed number of steps, unlike MCMC approaches.

More specifically, DSM can be phrased in terms of \gls*{sde}s \citep{song2021sde}, where training data is evolved into noise under a variance-exploding diffusion process $\dd \vb{x} = g(t) \, \dd \vb{w}$. Here $g(t)$ is called the diffusion coefficient, $\vb{w}$ is a Wiener process and $t \in [0, 1]$. The \gls*{sbm} learned using the DSM loss approximates the score $\grad_{\vb{x}} \log p_t(\vb{x})$ induced by this \gls*{sde}. The distribution $p_t(\vb{x})$ can be understood as the marginal distribution of trajectories from the \gls*{sde} evolved up to time $t$. Samples from the distribution $p(\vb{x}) = p_0(\vb{x})$ can then be generated by substituting this learned score into the corresponding reverse-time \gls*{sde} \citep{Anderson1982} $\dd \vb{x} = -g^2(t) \grad_{\vb{x}} \log p_t(\vb{x}) \dd t + g(t) \, \dd \bar{\vb{w}}$ and solving it with the distribution initialized to a wide Gaussian, $p_1(\vb{x}) = \normal(\vb{x} | \vb{0}, \sigma_\mathrm{max})$. Here $\bar{\vb{w}}$ is a reverse-time Wiener process and $\dd t$ is now an infinitesimal negative timestep. 

\subsection{Sampling from the posterior}
\label{sec:sampling-from-the-posterior}

\begin{figure}
    \centering
    \includegraphics[width=\linewidth]{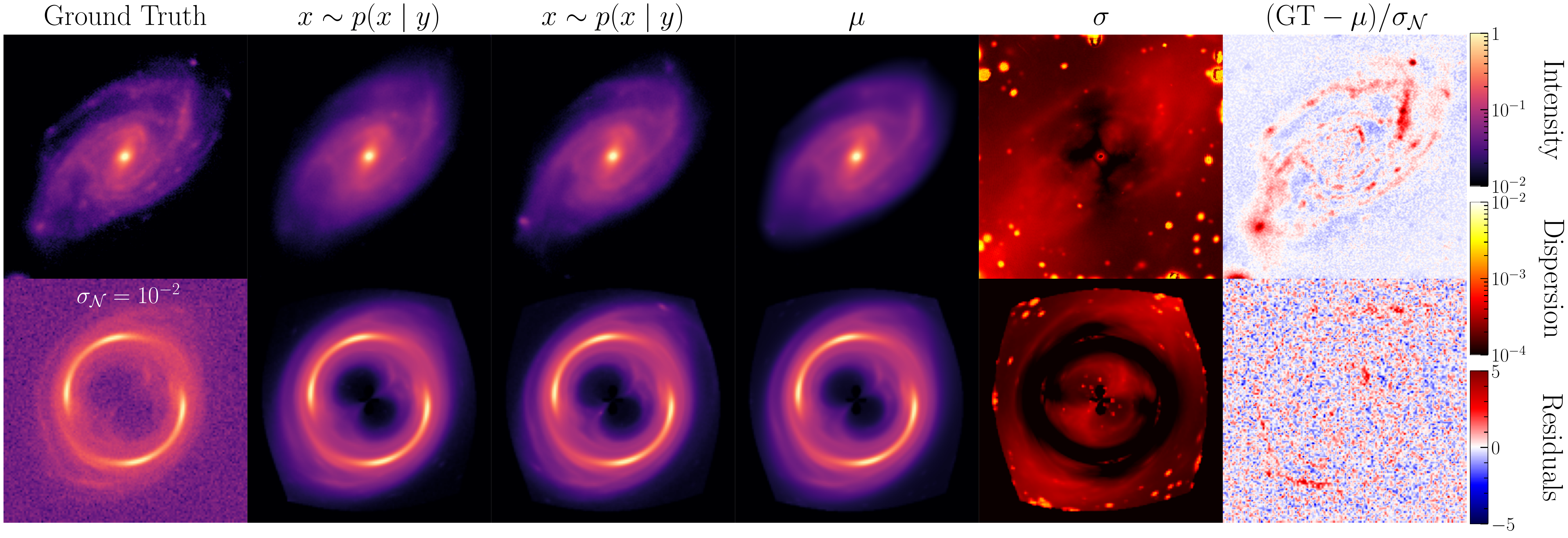}
    \caption{Source and observation reconstruction using \num{8000} steps of the Euler-Maruyama solver. The first column shows the true source image and the observation, labeled with the noise level $\noise$. The other columns, from left to right, show two samples from the posterior, the mean and standard deviation of \num{320} posterior samples, and the \gls*{snr} of the residuals in the source plane (first row) and lens plane (second row). The residuals in the source plane are shown for illustrative purposes only since the accuracy with which we expect the source to be reconstructed is position-dependent.}
    \label{fig:samples-mean-stdev}
\end{figure}

Sampling from the posterior $p(\vb{x} \mid \vb{y})$ requires changing the $t = 0$ boundary condition of the forward \gls*{sde} from the prior $p(\vb{x})$ to the posterior $p(\vb{x} \mid \vb{y})$. This modifies the reverse-time \gls*{sde} to read as \citep{song2021sde}
\begin{equation}
    \dd \vb{x} = -g^2(t) \grad_{\vb{x}} \log p_t(\vb{x} \mid \vb{y}) \dd t + g(t) \, \dd \bar{\vb{w}} \, .
\end{equation}
Solving this equation until $t=0$ then yields independent samples from $p(\vb{x} \mid \vb{y})$. We can further apply Bayes' rule to simplify the score in the above equation as
\begin{equation}
    \grad_{\vb{x}} \log p_t(\vb{x} \mid \vb{y}) = 
    \grad_{\vb{x}} \log p_t(\vb{x}) + \grad_{\vb{x}} \log p_t(\vb{y} \mid \vb{x}) \, .
\end{equation}
The first term in this sum is modeled by our \gls*{sbm}, $s_\theta(\mathbf{x}_t, t)$. We approximate the second using the \emph{convolved likelihood}, $p_t(\vb{y} \mid \vb{x}) \approx \normal(\vb{y} \mid A \vb{x}, (\noise^2 + \sigma^2(t)) \bbone)$.
Intuitively, this can be understood as arising from the convolution of the Gaussian diffusion $\mathcal{N}(\mathbf{0},\, \sigma^2(t)\bbone)$ with the Gaussian likelihood $p(\vb{y} \mid \vb{x})$. We derive the convolved likelihood in more detail in \cref{app:convolved-likelihood}. %

With this machinery in place, we can apply the Euler-Maruyama solver (see e.g.\ \citep{kloeden2011numerical}) to generate posterior samples of $\vb{x}$. The quality of the resulting samples is controlled by the number of time discretizations, $N$. We choose $N=8000$ based on Technique 2 of \citet{song2020improved}'s work, which we discuss in more detail in \cref{app:euler-maruyama}. This theory ensures that the samples from each iteration of the solver do not stray far from the high-density region of $p_t(\vb{x})$. The noise schedule in the DSM loss (or equivalently the diffusion coefficient in the forward \gls*{sde}) employed in this work corresponds to the \gls*{vesde} from \citet{song2021sde}. This amounts to setting $g(t) = \sqrt{\dv*{\sigma^2(t)}{t}}$ with $\sigma(t) = \sigma_{\mathrm{min}} \left( \sigma_{\mathrm{max}}/\sigma_{\mathrm{min}} \right)^{t}$. We explain how $\sigma_\mathrm{min/max}$ is selected in \cref{sec:results-conclusion}.

\section{Results and conclusions}
\label{sec:results-conclusion}

\begin{figure}
    \centering
    \resizebox{\linewidth}{!}{
        \begin{tikzpicture}
                \node at (0, 0) {\includegraphics[width=12cm]{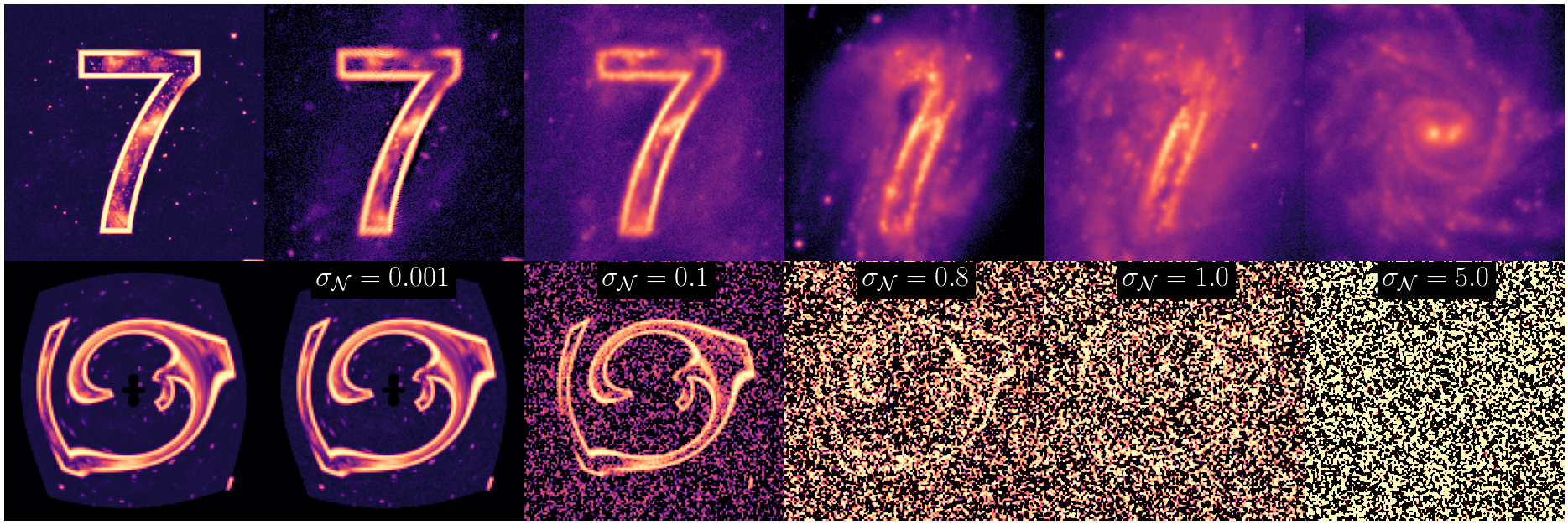}}; 
                \draw[-latex] (-3, -2.1) -- (5, -2.1); 
                \node at (0.5, -2.3) {\scriptsize \strut Data with increasing levels of noise};
                \node at (0.5, 2.2) {\scriptsize \strut Samples from the posterior $p(\mathbf{x} \mid \mathbf{y})$};
                \node at (-5, 2.2) {\scriptsize \strut Ground Truth};
                \node[rotate=90] at (-6.5, 1) {\scriptsize \strut Background };
                \node[rotate=90] at (-6.25, 1) {\scriptsize \strut source ($\mathbf{x}$)};
                \node[rotate=90] at (-6.5, -1) {\scriptsize \strut Distorted};
                \node[rotate=90] at (-6.25, -1) {\scriptsize \strut image ($\mathbf{y}$)};
        \end{tikzpicture}
    }
    \caption{Application of the method to a lensing system with a highly out-of-distribution source. The ground truth is given in the leftmost panel. Other panels show increasingly noisy data (lower row) and a sample from their corresponding source posterior (upper row). As the likelihood becomes less informative, the prior dominates, making the sources increasingly look like galaxies.}
    \label{fig:ood}
\end{figure}
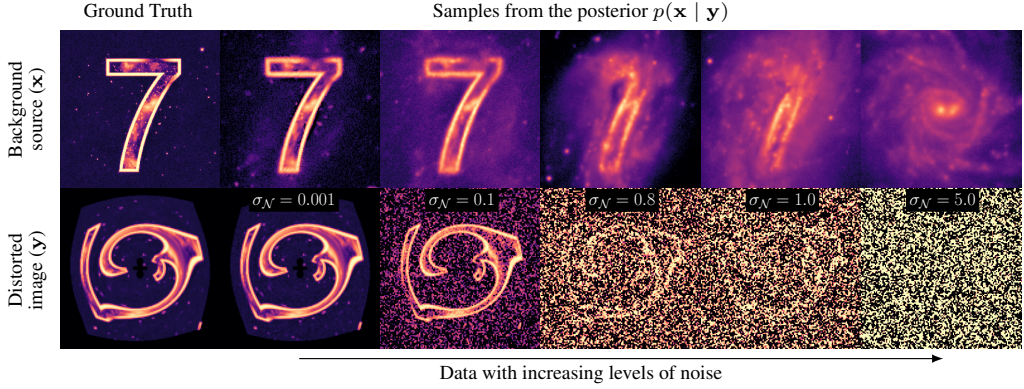

To test our method, we trained our \gls*{sbm} on the PROBES dataset \citep{2019ApJ...882....6S,2021ApJ...912...41S}, a high-quality sample of \num{2059} galaxies with $256\times256$ pixels. 
We used the data normalization scheme described in sec. 3.1 of \citet{astroddpm}. For the \gls*{sde} parameters we set $\sigma_\mathrm{max} = 263.4$ (our estimation of the largest Euclidean distance between any two pairs in our training set \citep{song2020improved}) and chose $\sigma_\mathrm{min} = 10^{-4}$, roughly the scale of the smallest details in the training images. Our network is the reference \texttt{PyTorch} \citep{pytorch} implementation of NCSN++ architecture from \citet{song2021sde}.\footnote{
    Available at \url{https://github.com/yang-song/score_sde_pytorch/}, released under Apache License Version 2.0.
} Training was carried out on four NVIDIA V100 GPUs, with a batch size of $16$, for a total of $\sim 350\,000$ optimization steps ($\sim 70$ hours wall-time). The results shown in \cref{fig:samples-mean-stdev} are in the $g$ band (though in principle our framework can be extended to multiband data). The lens deflections are modeled using a \gls*{sie} (see e.g.\ \citep{meneghetti2021introduction}) plus external shear. We produce noise-free images at $256\times256$ resolution by ray-tracing and bilinearly-interpolating the source over the deflected coordinates. We then pixelate at $128\times128$ resolution with average pooling. For sampling, we use 80 V100 GPUs in parallel, yielding \num{320} samples in less than one hour (wall-time).

In \cref{fig:samples-mean-stdev} we apply our method to a simulated lens from our test set. We show two posterior samples to give a sense of their variations, along with the mean and standard deviation calculated using \num{320} samples; see \cref{fig:post-samples} for more samples.  We find individual reconstructions and their mean match the observation almost down to the noise level. In the source plane, the bright core, spiral arms, and small but sharp clumps are well-reconstructed. Other small-scale features differ between the samples and have larger reconstruction uncertainties. The map of the standard deviation of the samples clearly shows fewer variations close to the diamond caustic, which matches our expectations, since these regions are highly magnified and thus better constrained. We find some posterior samples have bright peripheral spots, showing the model has learned they are present in the prior and that the approximation in our posterior sampling procedure does not suppress them.

Next, we apply our method to reconstructing an extremely out-of-distribution source.\footnote{
    Our source was generated with DALL-E 2 \citep{dalle2} using the prompt ``A galaxy in the shape of the number 7 on a dark background''. 
} The results are shown in \cref{fig:ood}. For low-noise observations, where the likelihood is highly informative, the model yields excellent reconstructions, capturing even small-scale spots in the source. 
This demonstrates that the model is qualitatively robust to distributional shifts when the likelihood is highly informative. With increasing noise levels the likelihood becomes less informative and the reconstructions increasingly resemble samples from the prior. As expected, the highly magnified regions in the center of the image (near the caustics) are better constrained. 

In conclusion, we combined a score-based model trained on images of real galaxies with a differentiable lensing likelihood to sample posteriors of pixelated sources in strong lenses. Our posterior samples have remarkably high fidelity to the ground truth, and our reconstructed observations are consistent with the true ones almost down to the noise level. The independent samples generated from the posterior allow us to assess the confidence of any features in the reconstructions (e.g., the existence of a spiral arm) by examining their variations in them. Through our experiment with out-of-distribution sources, we showed that our model can recover these sources when high-quality data make the likelihood informative and can converge to the learned prior when the likelihood is not constraining. We believe this inference approach will help enable new scientific analysis using existing and upcoming strong lensing observations.

\section*{Broader Impact}

The focus of this work is the rigorous estimation of uncertainties (posterior sampling) in high-dimensional spaces. The work can have an important cross-disciplinary impact on the application of machine learning in other natural sciences where accurate estimation of uncertainties is crucial. Given the striking nature of gravitational lensing images, we also believe that there is potential for a positive impact that will inspire broader interest in astrophysics.  While we do not anticipate our work could have direct negative consequences, it could conceivably be applied to ethically-questionable inference problems. Additionally, users of such methods must be aware of their approximations and biases when applying them to scientific problems.

\begin{ack}
    We are grateful for useful discussions with Pablo Lemos and Yatin Dandi.
    
    This research was made possible by a generous donation from Eric and Wendy Schmidt by recommendation of the Schmidt Futures Program. A.A.\ is supported through an IVADO Excellence Scholarship. Y.H.\ and L.P.L.\ acknowledge support from the National Sciences and Engineering Council of Canada Discovery grants RGPIN-2020-1505 and RGPIN-2020-05102, and also the Canada Research Chairs Program. Y.B.\ acknowledges funding from CIFAR, Samsung, IBM, Genentech, and Microsoft. This research was enabled in part by support provided by Calcul Québec (\url{https://www.calculquebec.ca/}) and the Digital Research Alliance of Canada (\url{https://alliancecan.ca/}). The Béluga cluster on which the computations were carried out is 100\% hydro-powered.

    Software used: \texttt{astropy} \citep{astropy:2013,astropy:2018}, \texttt{jupyter} \citep{jupyter}, \texttt{matplotlib} \citep{matplotlib}, \texttt{numpy} \citep{numpy}, \texttt{PyTorch} \citep{torch} and \texttt{tqdm} \citep{tqdm}.
\end{ack}

\bibliography{refs.bib}

\clearpage

 \section*{Checklist}
 \begin{enumerate}

 \item For all authors...
 \begin{enumerate}
   \item Do the main claims made in the abstract and introduction accurately reflect the paper's contributions and scope?
     \answerYes{}
   \item Did you describe the limitations of your work?
     \answerYes{We explain our convolved likelihood approximation in \cref{sec:sampling-from-the-posterior} and study its validity in \cref{app:convolved-likelihood}.}
   \item Did you discuss any potential negative societal impacts of your work?
     \answerYes{We address this in the "Broader Impacts" section.}
   \item Have you read the ethics review guidelines and ensured that your paper conforms to them?
     \answerNA{}
 \end{enumerate}

 \item If you are including theoretical results...
 \begin{enumerate}
   \item Did you state the full set of assumptions of all theoretical results?
     \answerYes{Please refer again to \cref{sec:sampling-from-the-posterior} and \cref{app:convolved-likelihood}}
   \item Did you include complete proofs of all theoretical results?
     \answerNA{}
 \end{enumerate}

 \item If you ran experiments...
 \begin{enumerate}
   \item Did you include the code, data, and instructions needed to reproduce the main experimental results (either in the supplemental material or as a URL)?
     \answerNo{}
   \item Did you specify all the training details (e.g., data splits, hyperparameters, how they were chosen)?
     \answerYes{We explain how we preprocess the data, set up our \gls*{sde} and train our networks in \cref{sec:results-conclusion}. Our networks and training setup is practically identical to the one in \citet{song2021sde}, with the exception of the smaller batch size, as mentioned in the text.}
   \item Did you report error bars (e.g., with respect to the random seed after running experiments multiple times)?
     \answerYes{Our main result is careful quantification of uncertainties when solving an inverse problem.}
   \item Did you include the total amount of compute and the type of resources used (e.g., type of GPUs, internal cluster, or cloud provider)?
     \answerYes{Please see \cref{sec:results-conclusion}. We acknowledge our computing cluster in the acknowledgements section.}
 \end{enumerate}

 \item If you are using existing assets (e.g., code, data, models) or curating/releasing new assets...
 \begin{enumerate}
   \item If your work uses existing assets, did you cite the creators?
     \answerYes{We reference the codebase from which we adapted our networks in \cref{sec:results-conclusion}.}
   \item Did you mention the license of the assets?
     \answerYes{Please refer to the footnote on page 3.}
   \item Did you include any new assets either in the supplemental material or as a URL?
     \answerNo{}
   \item Did you discuss whether and how consent was obtained from people whose data you're using/curating?
     \answerNA{}
   \item Did you discuss whether the data you are using/curating contains personally identifiable information or offensive content?
     \answerNA{}
 \end{enumerate}

 \item If you used crowdsourcing or conducted research with human subjects...
 \begin{enumerate}
   \item Did you include the full text of instructions given to participants and screenshots, if applicable?
     \answerNA{}
   \item Did you describe any potential participant risks, with links to Institutional Review Board (IRB) approvals, if applicable?
     \answerNA{}
   \item Did you include the estimated hourly wage paid to participants and the total amount spent on participant compensation?
     \answerNA{}
 \end{enumerate}

 \end{enumerate}

 \clearpage

\appendix

\section{The convolved likelihood}
\label{app:convolved-likelihood}

In this appendix, we explain the origin of the convolved likelihood approximation to $\log p_t(\vb{y} \mid \vb{x}_t)$ and demonstrate its regime of validity in the context of strong lensing source reconstruction. Let the Markov chain of the forward \gls*{vesde} be denoted as $\{ \vb{X}_t \}_{t \in [0, 1]}$. Our goal is to find a tractable expression for the marginal posterior $p_t(\vb{x}_t \mid \vb{y})$ at time $t$, the score of which is required to solve our reverse-time \gls*{sde}. By construction of the \gls*{vesde}, the random variable $\vb{X}_t = \vb{X}_0 + \vb{Z}_t$ can be expressed as the sum of the random variable $\vb{X}_0$, sampled from the posterior $\vb{x}_0 \sim p(\vb{x}_0 \mid \vb{y})$, and $\vb{Z}_t$, a noise perturbation $\vb{z}_t \sim \normal(\vb{0}, \Sigma_t)$, where $\Sigma_t := \sigma^2(t) \bbone$. This implies the marginal posterior we seek can be written as the convolution
\begin{equation}
    p_t(\vb{x}_t \mid \vb{y}) = \int \dd{\vb{x}_0} p(\vb{x}_0 \mid \vb{y}) \, \normal(\vb{x}_t \mid \vb{x}_0, \Sigma_t) \, .
\end{equation}
We can expand this expression by applying Bayes' rule to the first term in the integrand:
\begin{align}
    p_t(\vb{x}_t \mid \vb{y}) &= \frac{1}{p(\vb{y})} \int \dd{\vb{x}_0} p(\vb{y} \mid \vb{x}_0) \, p(\vb{x}_0) \, \normal(\vb{x}_t \mid \vb{x}_0, \Sigma_t) \\
    &= \frac{1}{p(\vb{y})} \int \dd{\vb{x}_0} \normal(\vb{y} \mid A \vb{x}_0, \Sigma_y) \, p(\vb{x}_0) \, \normal(\vb{x}_t \mid \vb{x}_0, \Sigma_t) \, , \label{eq:marginal-post-integral}
\end{align}
where we used the form of the lensing data generation process (\cref{eq:inverse problem}) to obtain the second line, with $\Sigma_y := \noise^2 \bbone$. Given a sufficiently broad prior $p(\vb{x}_0)$, the integral approximately factorizes into the product of the prior and likelihood convolved with the noise perturbation:
\begin{align}
    p_t(\vb{x}_t \mid \vb{y}) &\approx \frac{1}{p(\vb{y})} \left[ \int \dd{\vb{x}_0} \normal(\vb{y} \mid A \vb{x}_0, \Sigma_y) \, \normal(\vb{x}_t \mid \vb{x}_0, \Sigma_t) \right] \left[ \int \dd{\vb{x}_0} p(\vb{x}_0) \, \normal(\vb{x}_t \mid \vb{x}_0, \Sigma_t) \right] \\
    &= \frac{1}{p(\vb{y})} \left[ \int \dd{\vb{x}_0} \normal(\vb{y} \mid A \vb{x}_0, \Sigma_y) \, \normal(\vb{x}_t \mid \vb{x}_0, \Sigma_t) \right] p_t(\vb{x}_t) \\
    &= \frac{\normal(\vb{y} \mid A \vb{x}_t, \Sigma_y + A \Sigma_t A^T) \, p_t(\vb{x}_t)}{p(\vb{y})} \, , \label{eq:conv-likelihood-factorization}
\end{align}
where we applied the definition of $p_t(\vb{x}_t)$ to obtain the second equation and analytically evaluated the remaining integral (see e.g.\ sec. 3.3.1 of \citet{pml1Book} for the required identity). By expanding the left-hand side of this equation with Bayes' rule, we obtain the convolved likelihood,
\begin{equation}
    p_t(\vb{y} \mid \vb{x}_t) \approx \normal(\vb{y} \mid A \vb{x}_t, \Sigma_y + A \Sigma_t A^T) \, .
\label{eqn12}
\end{equation}

We can examine the accuracy of the convolved likelihood factorization by considering the case where $p(\vb{x}_0) = \normal(\vb{x}_0 \mid \vb{0}, \Sigma_{x_0})$. In this case the integral in \cref{eq:marginal-post-integral} giving the marginal posterior can be evaluated analytically:
\begin{align}
    p_t(\vb{x}_t \mid \vb{y}) &= \frac{1}{p(\vb{y})} \int \dd{\vb{x}_0} p(\vb{y} \mid \vb{x}_0) \, p(\vb{x}_0) \, p(\vb{x}_t \mid \vb{x}_0) \\
    &= \frac{1}{p(\vb{y})} \int \dd{\vb{x}_0} \normal(\vb{y} \mid A \vb{x}_0, \Sigma_y) \, \normal(\vb{x}_0 \mid \vb{0}, \Sigma_{x_0}) \, \normal(\vb{x}_t \mid \vb{x}_0, \Sigma_t) \\
    &= \frac{\normal(\vb{x}_t \mid \vb{0}, \Sigma_{x_0} + \Sigma_t)}{p(\vb{y})} \int \dd{\vb{x}_0} \normal(\vb{y} \mid A \vb{x}_0, \Sigma_y) \, \normal(\vb{x}_0 \mid \vb{m}_c, \Sigma_c) \\ \label{eq:intermediate-step}
    &= \frac{\normal(\vb{x}_t \mid \vb{0}, \Sigma_{x_0} + \Sigma_t) \, \normal(\vb{y} \mid A \vb{m}_c, \Sigma_y + A \Sigma_c A^T)}{p(\vb{y})} \, ,
\end{align}
where we obtained \cref{eq:intermediate-step} by simplifying the product of the last two terms in the integrand using eq. 371 from \citet{petersen2008matrix} and defined
\begin{equation*}
    \Sigma_c := (\Sigma_{x_0}^{-1} + \Sigma_t^{-1})^{-1} \, , \qq{} \vb{m}_c := \Sigma_c \Sigma_t^{-1} \vb{x}_t \, .
\end{equation*}
On the other hand, evaluating the integral $p_t(\vb{x})$ in our convolved likelihood factorization \cref{eq:conv-likelihood-factorization} yields
\begin{align}
    p_t(\vb{x}_t \mid \vb{y}) &\approx \frac{\normal(\vb{x}_t \mid \vb{0}, \Sigma_{x_0} + \Sigma_t) \, \normal(\vb{y} \mid A \vb{x}_t, \Sigma_y + A \Sigma_t A^T)}{p(\vb{y})} \, .
\end{align}

Thus, our approximation for the convolved likelihood holds if ${\normal(\vb{y} \mid A \vb{m}_c, \Sigma_y + A \Sigma_c A^T) \approx \normal(\vb{y} \mid A\vb{x}_t, \Sigma_y + A \Sigma_t A^T)}$. Such an approximation is valid when the prior $p(\vb{x}_0)$ is broad compared to the likelihood. To test this, we expand $\Sigma_c$ in a Neumann series around $\Sigma_t$:
\begin{equation}
    \Sigma_c = \Sigma_t - \Sigma_t \Sigma_{x_0}^{-1}\Sigma_t + \mathcal{O}(\lVert  \Sigma_t (\Sigma_{x_0}^{-1} \Sigma_t)^2 \rVert_{2}) \, ,
\end{equation}
where $\lVert \cdot \rVert_2$ is the spectral norm of a matrix, i.e.~the magnitude of the largest eigenvalues of the covariance matrix. Our approximation holds when the second and higher order terms are negligible compared to the leading term in the expansion. We estimated the eigenvalues of $\Sigma_{x_0}$ by fitting a Gaussian random field on the PROBES dataset. Our estimate of the largest eigenvalue of $\Sigma_{x_0}$ is comparable to $\sigma_{\mathrm{max}}^2$, which means that our approximation might not hold for $t \lesssim 1$, but that it will be valid for most of the sampling procedure $0 \leq t < 1$. Moreover, it is worth noting that in the limit $t\rightarrow 0$, the approximation in eq. (\ref{eqn12}) becomes exact, and so the approximate SDE we are solving respects the same boundary condition as the exact SDE.

A final simplification we apply to our convolved likelihood to avoid an expensive matrix inversion at every step while solving the reverse-time \gls*{sde} is to assume $A A^T \approx \bbone$. Without this assumption, evaluating the convolved likelihood would require inverting the matrix $\Sigma_y + A \Sigma_t A^T$, the size of which is the number of pixels in the image squared. With this assumption, on the other hand, the covariance matrix of the convolved likelihood simplifies to $\Sigma_y + \Sigma_t \propto \bbone$, which is trivial to invert. By examining $A A^T$ for different lens configurations, we find this is a reasonable approximation. In our checks, only a small fraction of off-diagonal elements are nonzero, and all diagonal elements are guaranteed to be no greater than $1$ since lensing conserves surface brightness.\footnote{
    Note that some rows and columns of $A A^T$ may contain only zeros. These rows/columns correspond to pixels in the image $\vb{y}$ that trace back to points in the source plane outside of the region where the pixelated source $\vb{x}$ is defined. Such pixels, therefore, have no impact on the source reconstruction and can be ignored.
} Consequentially, the final convolved likelihood we use for sampling is
\begin{equation}
    p_t(\vb{y} \mid \vb{x}_t) \approx \normal(\vb{y} \mid A \vb{x}_t, \Sigma_y + \Sigma_t) \, .
\end{equation}

\section{Euler-Maruyama discretization}
\label{app:euler-maruyama}

The Euler-Maruyama discretization of the reverse-time SDE is
\begin{equation}\label{eq:notes euler maruyam}
        \mathbf{x}_{t+\Delta t} = \mathbf{x}_{t} - g^2(t ) \grad_{\mathbf{x}_{t}} \log p_{t}(\mathbf{x}_{t} \mid \mathbf{y}) \Delta t + g(t) \mathbf{z}_{t}\sqrt{-\Delta t}  
\end{equation} 
with $\mathbf{z}_t \sim \mathcal{N}(\mathbf{0}, \bbone)$, $\Delta t = -1/N$ and $N$ the number of discretisations of the time index $t \in [0, 1]$. In practice, we can choose $N$ to satisfy technique 2 of \citet{song2020improved}, such that the discretized noise schedule used in our work $\sigma(t) = (\sigma_{\mathrm{max}} / \sigma_{\mathrm{min}})^t \sigma_{\mathrm{min}}$ is now a geometric progression with a ratio
\begin{equation}\label{eq:notes ratio}
        \gamma = \frac{\sigma(t)}{\sigma(t+\Delta t)} = \left( \frac{\sigma_{\mathrm{max}}}{\sigma_{\mathrm{min}}} \right)^{1/N}\, .
\end{equation}
The ratio $\gamma > 1$ should be close enough to $1$ so that a sample from $p_{t}(\vb{x}_t)$ should at least belong to the $3\sigma$ density region of $p_{t + \Delta t}(\vb{x}_t)$. In such a situation, a sample from $p_{t}(\vb{x}_t)$ will have some probability $p(N)$ of belonging to the $3\sigma$ density region of $p_{t + \Delta t}(\vb{x}_t)$, meaning it is a likely sample even after a transition to the density distribution at lower temperature.

We follow \citet{song2020improved} in setting this probability to
\begin{equation}\label{eq:notes p transition}
        p(N) = \Phi(\sqrt{2n}(\gamma - 1) + 3\gamma) - \Phi(\sqrt{2n}(\gamma - 1) - 3\gamma)
\end{equation}
where $\Phi$ is the CDF of a normal distribution. 
For a stable diffusion, we ask that $p(N) \gtrsim 0.5$. For the dimensionality of our problem ($n=256^2$ for the PROBES dataset), we thus have that $N=2000$ minimally satisfy this criteria with $p(2000) = 0.64$. We can increase our confidence in the solver by setting $N=8000$, s.t. $p(8000) = 0.99$, which is what is used in this work. 


\newpage
\clearpage

\section*{Additional figures}

\begin{figure}[h]
    \centering
    \includegraphics[width=\textwidth]{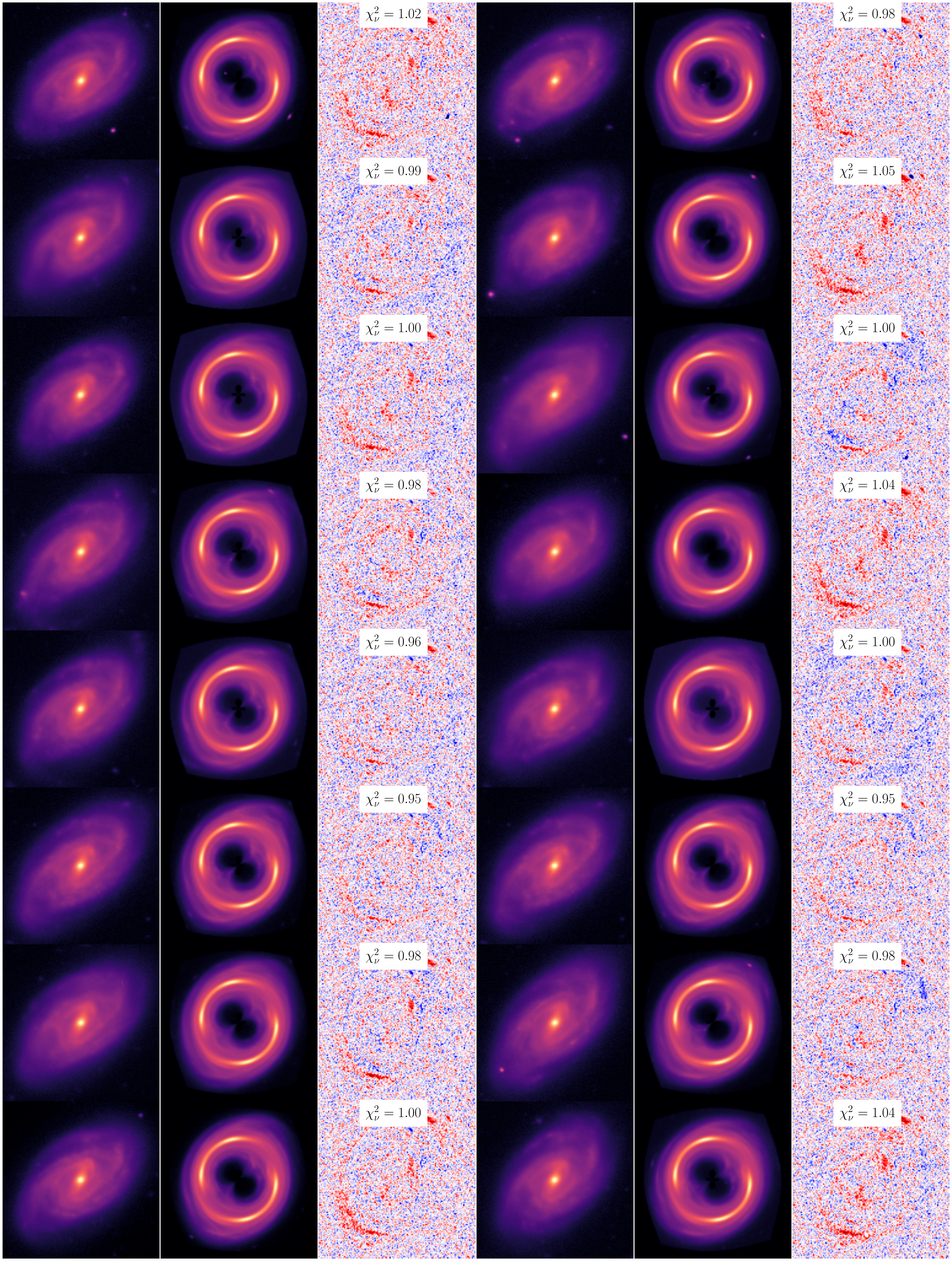}
    \caption{Additional posterior samples for the observation from \cref{fig:samples-mean-stdev} in the source and lens planes, along with the $\chi^2_\nu$ of the residuals in the lens plane. The noise 
    level in the data is $\sigma_{\mathcal{N}} = 0.01$.}
    \label{fig:post-samples}
\end{figure}

\end{document}